\documentclass[sigconf]{acmart}

\setcopyright{none}
\renewcommand\footnotetextcopyrightpermission[1]{} % removes footnote with conference information in first column
\AtBeginDocument{%
  \providecommand\BibTeX{{%
    \normalfont B\kern-0.5em{\scshape i\kern-0.25em b}\kern-0.8em\TeX}}}

\usepackage{afterpage}

\begin{document}

\title{Cinematic Techniques in Narrative Visualization}

% \author{Anonymous for Review}
\author{Matthew Conlen}
\affiliation{%
  \institution{Our World in Data}
}
\email{matt.conlen@ourworldindata.org}

\author{Jeffrey Heer}
\affiliation{%
  \institution{University of Washington}
  }
\email{jheer@uw.edu}

\author{Hillary Mushkin}
\affiliation{%
  \institution{California Institute of Technology}}
\email{hmushkin@caltech.edu}

\author{Scott Davidoff}
\affiliation{%
  \institution{Jet Propulsion Laboratory\\California Institute of Technology}}
\email{scott.davidoff@jpl.nasa.gov}

\renewcommand{\shortauthors}{Conlen, et al.}

\begin{abstract}
% NASA’s Earth Science Communications team uses narrative visualizations to help millions of followers connect to climate change data. To convey the urgency of climate issues, we developed cinematic visualizations that help people understand planetary-scale data on a human scale. 

The many genres of narrative visualization (e.g. data comics, data videos) each offer a unique set of affordances and constraints. To better understand a genre that we call \textit{cinematic visualizations}---3D visualizations that make highly deliberate use of a camera to convey a narrative---we gathered 50 examples and
% that fit our 
% utilizing one or more cameras to observe data embedded in a three-dimensional, time varying scene. We analyze the and 
analyzed their traditional cinematic aspects 
% these visualizations
to identify the benefits and limitations of the form. While the cinematic visualization approach can violate traditional rules of visualization, we find that through careful control of the camera, cinematic visualizations enable immersion in data-driven, anthropocentric environments, and can naturally incorporate in-situ narrators, concrete scales, and visual analogies. Our analysis guides our design of a series of cinematic visualizations, created for NASA’s Earth Science Communications team. We present one as a case study to convey design guidelines covering cinematography, lighting, set design, and sound, and discuss challenges in creating cinematic visualizations. 
\end{abstract}

\maketitle

\newcommand{\todo}[1]{{\textcolor{red}{[TODO: #1]}\normalfont}}
\newcommand{\matt}[1]{{\textcolor{blue}{[MC: #1]}\normalfont}}
\newcommand{\scott}[1]{{\textcolor{orange}{[SD: #1]}\normalfont}}
\newcommand{\hillary}[1]{{\textcolor{purple}{[HM: #1]}\normalfont}}
\newcommand{\jeff}[1]{{\textcolor{olive}{[JH: #1]}\normalfont}}
\newcommand{\note}[1]{{\textcolor{red}{[Note: #1]}\normalfont}}

\newcommand{\CV}[1]{{\textcolor{red}{#1}\normalfont}}

\section{Introduction}

Within narrative visualization~\cite{segel2010narrative}, researchers have identified genres (such as  data comics~\cite{bach2018design} and data videos~\cite{amini2015understanding}) that help better unpack and situate their specific application and the features that they employ. 
% We observe an outlier class whose characteristics are not encapsulated by the current genre classifications. We call these 
\textbf{cinematic visualizations}
% : narrative visualizations that 
embed data into a three-dimensional, time varying scene, utilizing one or more cameras to direct the relationship between a viewer and the scene to tell a dramatic data-driven story. 
% We contend that this growing body of work represents an emergent genre of narrative visualization.
% They draw on techniques from digital animation, film, and scientific visualization. 
This cinematic approach is different from the one typically used in information visualization, where graphics are reduced to a minimal form, incorporating only essential elements like axes and data-driven marks~\cite{tufte2001visual}. Cinematic visualizations are more maximal: non-data marks are not compressed or reduced, instead entire digital worlds are built up around data points and included in the visible frame.  
This technique allows viewers to feel present in locations augmented with data-bound objects, known as data visceralizations~\cite{lee2020data}. 
Narrative documentary visualizations~\cite{bradbury2020documentary} can be produced through the careful editorial direction of the  \textit{cinematography}, \textit{editing}, \textit{mise-en-sc\`ene}, and \textit{sound}~\cite{bordwell1997film}. 

Through an analysis of 50 existing cinematic visualizations, we identified four salient techniques (in-situ narrators, resolution of scale, anthropocentric perspective,  and story-driven cameras) that cinematic visualizations employ to dramatically engage their audience through emotionally resonant data-stories.
% These visualizations sometimes include on-screen narrators who interact directly with the data visceralizations (\textit{in-situ narrators}).
% These graphics are becoming popular among practitioners but often break conventions about when to use 3D in visualization. 
% We collected 50 cinematic visualizations and analyzed them using Bordwell \& Thompson's formal system of film style~\cite{bordwell1997film} to understand why the designers of these graphics strayed from information visualization convention. 
% Through our analysis we identified four salient techniques (in-situ narrators, resolution of scale, anthropocentric perspective,  and interactive cameras) that cinematic visualizations employ to dramatically engage their audiences. 
We show how these techniques are used throughout the examples analyzed,  discuss constraints associated with them, 
% We identify narrative use cases where cinematic visualization has been applied, and
and reason about why cinematic visualizations may be effective despite the known pitfalls of 3D visualization. 
% We discuss how these pitfalls are (or are not) dealt with, and discuss why cinematic visualizations seem well suited for conveying scale, scientific storytelling, and (re-)enacting events.

Using the lessons learned from this formal analysis, we produced a web-based article 
% \textit{Visualizing the Quantities of Climate Change}, 
containing a series of cinematic visualizations relating to climate change, which was published by NASA's Earth Science Communications team~\footnote{\url{https://climate.nasa.gov/news/2933/visualizing-the-quantities-of-climate-change/}}. %https://perma.cc/KE76-MGJY}}
We contribute the design process for one of these visualizations as a case study, presenting design artifacts that were created during our process (both successful and unsuccessful), and provide concrete guidelines for designers of cinematic visualizations. 
Our analysis and design artifacts are available at \url{https://cinematic-visualization.github.io/}.

% or to experience events that happened in the past, or are happening far away in the universe.

\section{Related Work}

% NASA communications are developed primarily for the citizens of the United States. The Yale Program on Climate Change Communication has conducted an extensive audience survey and identified ``Global Warming’s Six Americas,'' six unique American audiences that model attitude toward climate change.

% We begin by exploring the emergence of Narrative Visualization.
Narrative visualizations are used to improve memorability~\cite{borkin2015beyond, borkin2013makes}, to instill empathy or emotion~\cite{boy2017showing}, to frame a message~\cite{hullman2011visualization}, and to improve engagement~\cite{greussing2019simply,2019-idyll-analytics}. Segel \& Heer~\cite{segel2010narrative} provided an initial characterization of the design space of narrative visualizations, which was later elaborated to include additional  techniques~\cite{stolper2016emerging}. Hullman et al.~\cite{hullman2013deeper} focused on the role of sequence in narrative visualization, characterizing a set of transition types and other high level strategies for sequencing visualizations. Tools have been created to support narrative visualization authoring~\cite{satyanarayan2014authoring,2018-idyll,TimelineStoryteller, amini2016authoring}, and a small number of empirical evaluations of narrative visualizations have been conducted~\cite{mckenna2017visual, zhi2019linking, 2019-idyll-analytics, boy2017showing}. Further work has investigated specific genres of narrative visualization such as data comics~\cite{bach2018design}, and new genres have emerged beyond Segel \& Heer's initial set, such as ``scrollytelling.'' 
 Here we add to the ongoing conversation around narrative visualization by identifying another such genre: \textit{cinematic visualization}. 
Kosara \& McKinley~\cite{kosara2013storytelling} identified the opportunity for narrative visualization researchers to learn from other disciplines that engaged heavily with storytelling and multimedia, this paper draws on film art scholarship, incorporates a formal system of cinematic style into our analysis, discussion, and design of cinematic visualizations.  

\begin{figure*}[ht]
  \centering
  \includegraphics[width=\linewidth]{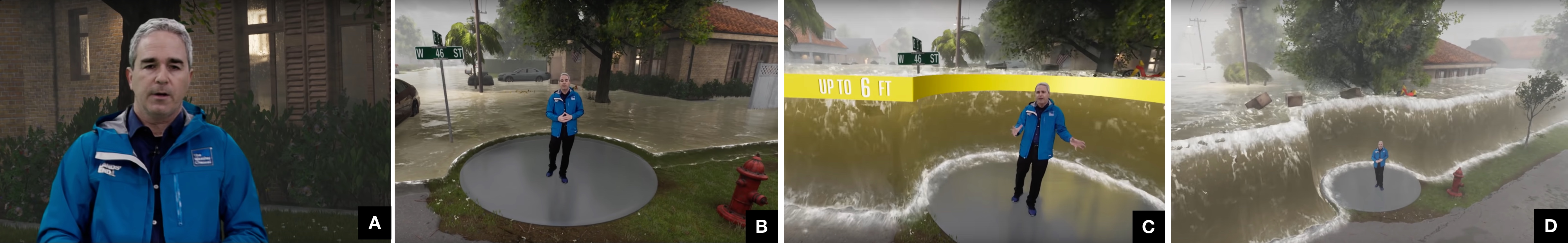}
  \vspace{-18pt}
  \caption{\textit{The Dangers of Storm Surge} (CV42) is a mixed reality video produced by the Weather Channel. 
  The video opens with a close up shot of a news anchor wearing a rain jacket, standing in front of a house (\ref{fig:weather}A). There are audible sounds of rain under the anchor's voice and water dripping down the windows of the house. The camera pulls back revealing that the live anchor is being composited into a 3D scene of a suburban neighborhood during a storm surge (\ref{fig:weather}B). There are very few data points actually encoded as visual elements. The piece simply shows water rising from zero, to three, to six, to nine feet (\ref{fig:weather}C-D) as the anchor narrates with details in reference to the danger of storm surge associated with hurricanes. 
%   The neighborhood backdrop helps contextualize the real-world impact that this event would have on a viewer; the items placed in the neighborhood serve as reference points, helping to turn an abstract number like ``6 feet of water'' into a concrete concept. As the water rises, it overtakes key objects which serve as a concrete reference for the viewer: first a car, then the narrator, then a street sign, then eventually to the windows of the house from the initial shot.
}
  \label{fig:weather}
%   \vspace{-20pt}
\end{figure*}

% Several tools have been developed to support the creation of a broad range of narrative visualizations, for example Ellipsis~\cite{satyanarayan2014authoring}, Idyll~\cite{2018-idyll} and TimelineStoryteller~\cite{TimelineStoryteller}. 

 \subsection{Data Videos \& VR}
 \textit{Data videos} were included in the initial set of genres put forth by Segel \& Heer~\cite{segel2010narrative} and first studied closely by Amini et al.~\cite{amini2015understanding}. Not all data videos are cinematic visualizations (for example, we do not consider a video consisting of a sequence of two-dimensional infographics to be cinematic), and not all cinematic visualizations are data videos (for example one in which the visualization is deeply tied to the text of an interactive news article). While Amini et al. were primarily concerned with the narrative structure and attention cues of data videos, we additionally consider the visual and auditory style of cinematic visualizations in detail.
%  , and identify visual and auditory techniques with applications to narrative visualization. 
 Under our formal style system, our analysis of \textit{editing} is most closely related to Amini's work, however that is only one of four dimensions we consider.
 
 Bradbury \& Guadagno~\cite{bradbury2020documentary} studied viewer preferences in \textit{documentary narrative visualization} (a subgenre of data videos in which data is presented using the techniques of documentary film), and found that audiences may prefer when documentary data videos include voice-over narration and on-screen narrators. We build on their analysis of the use of narrators and narration, in particular during our discussions of \textit{in-situ narrators} that interact with data-bound objects digitally rendered into the space around them, and of the use of sound in cinematic visualizations. Video producers have extended the traditional documentary visualization format to enable interactivity such as user selected paths through the content and manipulable graphics~\cite{fallenww2, virusbeautybeast}.

%  and created DataClips~\cite{amini2016authoring}, a narrative visualization authoring tool focused specifically on data videos. 

%  In both authoring tools and prior analyses, work on data videos has considered mostly the sequence and timing of juxtaposed audio, text, and 2D visualizations. We consider particularly those that use 3D graphics, and the challenges and (possible) benefits associated.

Immersive data stories~\cite{isenberg2018immersive} have been discussed within the emerging field of immersive analytics~\cite{marriott2018immersive} and have been shown to allow viewers to examine data at multiple scales, support immersive exploration, and create affective personal experiences with data~\cite{ivanov2019walk}. Lee et al.~\cite{lee2020data} introduced data visceralizations, where physical quantities are visualized in 3D virtual reality scenes. This paper helps to bridge the gap between data visceralization and narrative visualization by showing how cinematic techniques can be used to create author-guided narrative visualizations using data visceralizations. Cinematic visualizations similarly attempt to immerse viewers and create emotionally resonant experiences, although in contrast to immersive visualizations they are typically viewed on a standard 2D screen with limited (or no) user control of the camera. There are several toolkits for creating immersive data visualizations and data stories on augmented reality devices ~\cite{sicat2018dxr, cordeil2019iatk, ren2019visualization}.

\begin{figure*}[ht!]
  \centering
  \includegraphics[width=\linewidth]{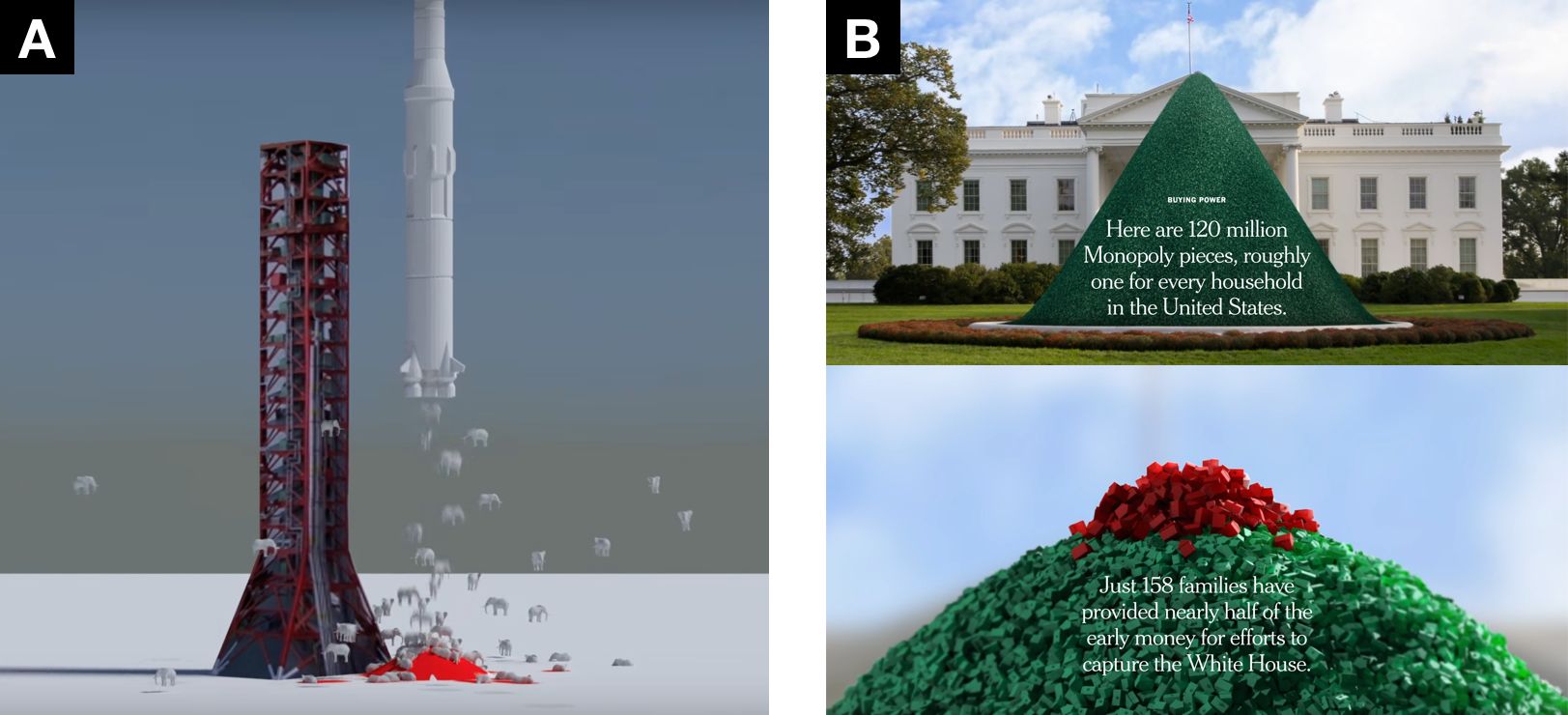}
%   \vspace{-18pt} 
  \caption{
%   Cinematic visualizations can naturally express techniques like visual analogies, unit visualization, and concrete scales to facilitate comparison and allow viewers to understand quantities of scale. 
  (A) In \textit{[REALISTIC] Elephant rocket fuel - Saturn V} (CV29), a model Saturn V rocket takes off, however, instead of flames exiting the bottom of the spacecraft, elephants are expelled, the number of elephants represents the corresponding mass of fuel. 
%   The visualization takes a humorous tone, as the out-of-place elephants bounce off the ground and techno music plays in the background. 
  This video may not make for a particularly effective visualization in terms of conveying precise quantities, but the style successfully uses humor in order to call attention to the fact that rocket launches use a quantity of fuel so great it is appropriate to measure it in terms of dozens of elephants. (B) In \textit{Here are 120 million Monopoly pieces, roughly one for every household in the United States} (CV6) by the New York Times the pile of Monopoly pieces is first seen from a far, before the reader scrolls down the page to trigger the camera zooming in to the very top of the pile, dramatically revealing what a disproportionately small portion of families provide most political funding.}
  \label{fig:scale}
  \vspace{-10pt}
\end{figure*}

\subsection{3D Computer Graphics}

Animation~\cite{thompson2020understanding} has been a partner discipline with visualization for some time. Classic principles of animation~\cite{thomas1995illusion} have been adapted for digital usage~\cite{lasseter1987principles} and subsequently for information visualization~\cite{heer2007anim}. With realistic camera models~\cite{kolb1995realistic} and improving rendering capabilities~\cite{cook1987reyes} digital animation became a tool to create  Hollywood films~\cite{henne1996making}. While 3D graphics have been used in visualization to limited success, e.g., to display hierarchical information~\cite{cockburn2000evaluation}, the use of 3D graphics in information visualization is often avoided. A broad body of research documents potential pitfalls, including that volume is not a perceptually effective encoding channel~\cite{cleveland1984graphical}, and that 3D projections introduce distortion and occlusion \cite{ware2012information}.  We find that designers of cinematic visualizations may intentionally use suboptimal encodings in support of more visceral~\cite{lee2020data} and emotionally resonant~\cite{greussing2019simply} graphics. 
% Tools like Blender~\cite{blender} are used by artists to create animated 3D renderings. Improved access to such tools may be associated with the increasing number of cinematic visualizations that have been produced. 
% \matt{Add comment about how these tools are insuficient.}
 
The use of 3D does find more regular application in scientific visualization~\cite{upson1989application, baker1995after}, including its use in storytelling~\cite{ma2011scientific, cox2006metaphoric}. Borkiewicz used the term \textit{cinematic scientific visualization} ~\cite{Borkiewicz:2019:CSV:3305366.3328056}  to refer to a class of narrative data videos that focus on scientific data. Here we use \textit{cinematic visualization} in a similar way but do not restrict the data to be strictly scientific or inherently spatial. 
Unlike Borkiewicz, our description encapsulates visualizations which are not embedded in films, but may be, for example, displayed as an animation accompanying a news article.

% \todo{Add relevant citations of cinematic techniques from scientific visualization community, e.g. automated path planning for cameras.}

\subsection{Film Art}

% There is a long history of the criticism of works of cinema. 
% We bring this body of work into our research, 
% as a contribution, 
% relying on its formal language as a lens that we use to unpack the idea of a cinematic visualization. 
Bordwell and Thompson~\cite{bordwell1997film} define \textit{narrative} and \textit{style} as the two major formal systems of film. While prior work has examined sequence~\cite{hullman2013deeper} and narrative structure \& attention cues~\cite{amini2015understanding} in data videos, we observe that cinematic style has far less visibility in the critical vocabulary of data visualization. Style plays a crucial roll in filmmaking, enabling directors to ``confirm our expectations, or modify them, or cheat, or challenge them.  [...] A director directs not only the cast and crew. A director also directs us, directs our attention, shapes our reaction.''~\cite{bordwell1997film} This paper brings Bordwell and Thompson's formal system of cinematic style into the world of data visualization, and uses it to examine how narrative visualizations borrow techniques from cinema while departing from many of the traditional practices advocated by visualization research. 

Style consists of four features, which together make up a film's style, each now briefly described. \textit{Mise-en-sc\`ene} refers to everything that is seen in the frame, including lighting, actors, objects, backdrops, and so on~\cite{gibbs2002mise}. \textit{Cinematography} refers to the use of the camera, how shots are composed and framed~\cite{freytag1904technique}. 
% Photographic rules (e.g. the ``rule of thirds''~\cite{freeman2007photographer}) tell us that the locations of elements within shots will determine how the frame is decoded. 
By placing elements at specific locations within the frame, they can be perceived either as the subject or the background of the image~\cite{freeman2007photographer}.  \textit{Editing} is the composition of multiple pieces of footage in time or space, creating transitions between perspectives and scenes~\cite{reisz1971technique}. \textit{Sound} is the audio used, whether it be music, voice over, or sounds from characters or objects on screen~\cite{holman2012sound}. Our analysis of cinematic visualization identified techniques along these dimensions of style that designers can use to enhance their presentation of data narratives.

% It is not only the narrative system, but also the style that affects how a viewer experiences a film. 

% Style is important for its ability to impact a viewer's experience.  

\begin{figure}
  \includegraphics[width=\linewidth]{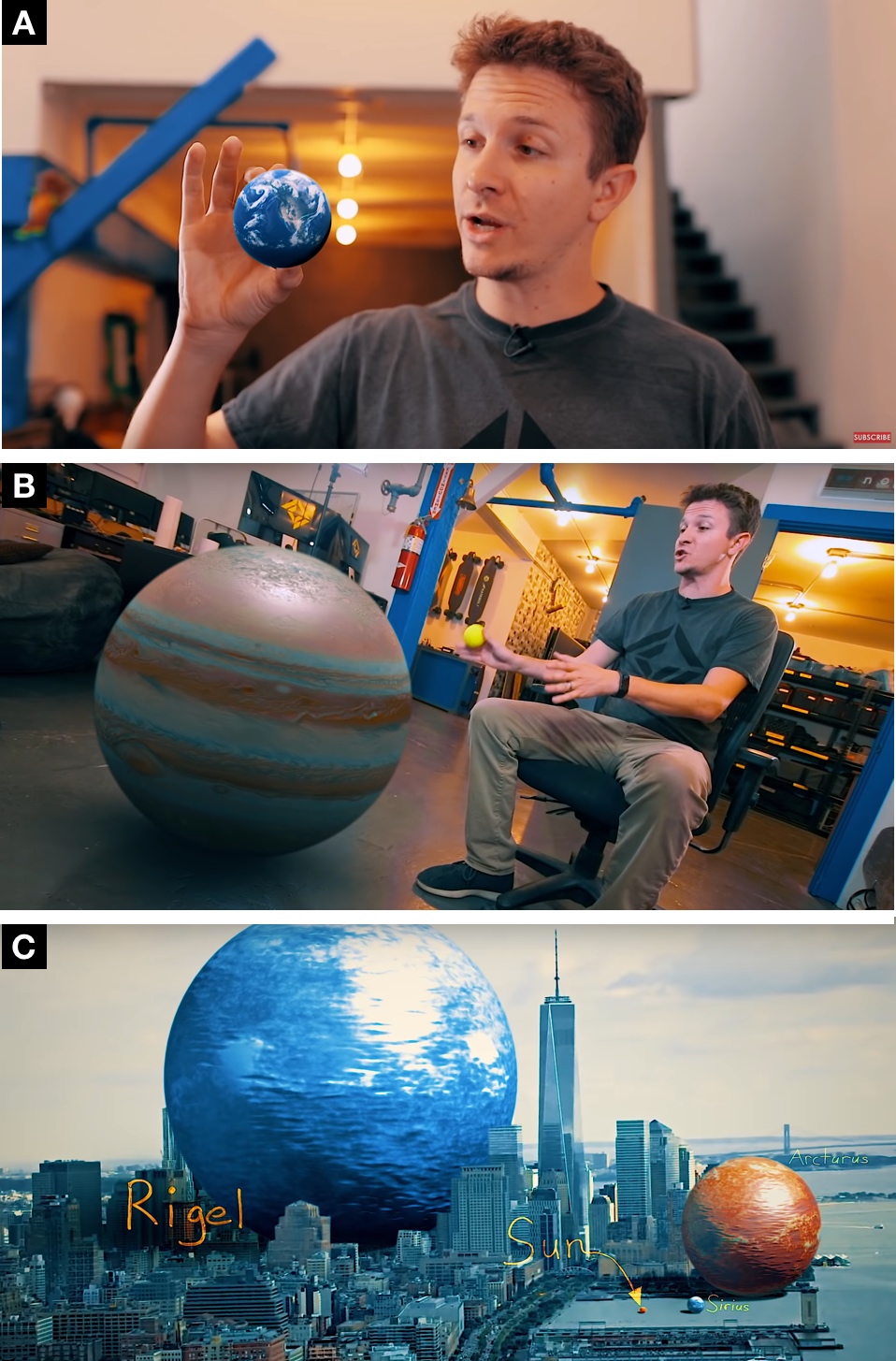}
%   \vspace{-18pt}
  \caption{\textit{VFX Artist Reveals the True Scale of the Universe} features a live-action narrator alongside scaled-down 3D models of celestial bodies.}
  \label{fig:mixed-reality}
%   \vspace{-15pt}
\end{figure}

% In the remainder of this paper, we unpack how Bordwell and Thompson's formal system for analyzing style can be applied to a data visualization to bring about a cinematic way of presenting a data narrative. Through these four features we analyze the style of 49 cinematic visualizations. Using this system we assess how cinematic visualizations use visual and auditory techniques to build a narrative based on data while directing an audience's attention and reaction to the story.  

\section{Cinematic Visualization Survey}

% \subsection{Data Collection \& Analysis}

% While reviewing existing narrative visualizations to inform our design of new graphics for NASA's climate change site, we identified cinematic visualizations, an emerging style that was not captured by existing genres put forth by Segel \& Heer and others. 

\begin{figure*}
  \centering
  \includegraphics[width=\linewidth]{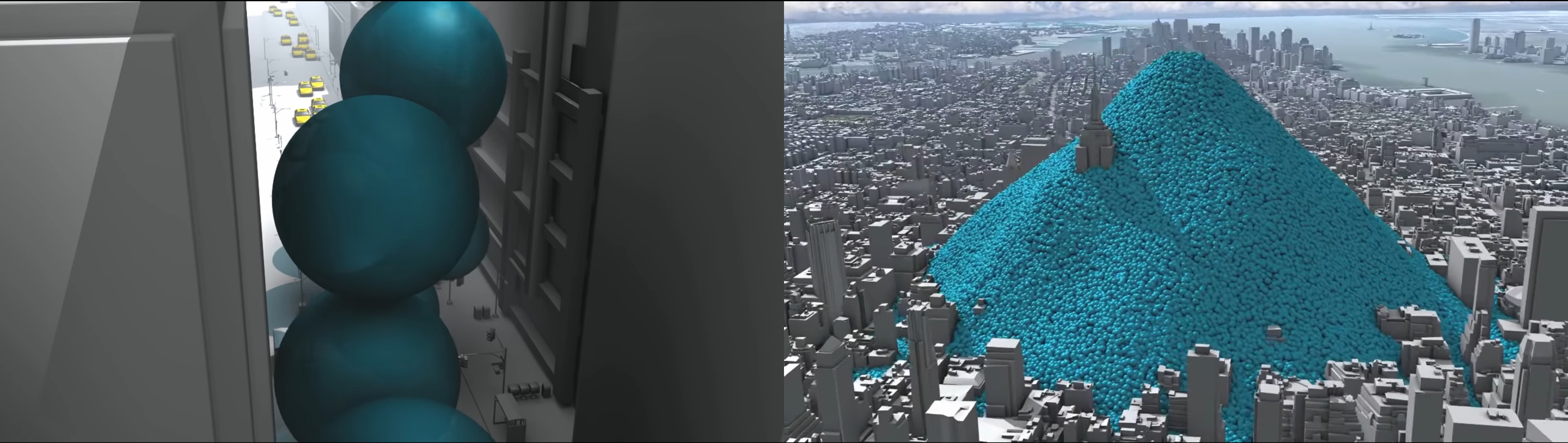}
  \caption{\textit{New York City's greenhouse gas emissions as one-ton spheres of carbon dioxide gas}, a cinematic visualization produced by Carbon Visuals and released online. The cinematic visualization uses a variety of different camera views, along with stark colors to guide viewers through an explanation of the scale of the city's greenhouse gas emissions. The number of instances of the blue sphere is driven by the rate of emissions. As this number grows the city buildings serve as a concrete scale.}
  \label{fig:nyc}
\end{figure*}

We collected cinematic visualization to analyze by surveying literature on narrative visualization~\cite{segel2010narrative, stolper2016emerging, isenberg2018immersive, ma2011scientific, Borkiewicz:2019:CSV:3305366.3328056, hullman2013deeper, amini2015understanding}, browsing the information visualization awards website Information is Beautiful~\cite{mccandless2012information} and the PacificVIS storytelling contest~\cite{pacvisstorytelling}, and searching for news articles, blog posts, conference talks, and videos which were described using combinations of the keywords \textit{cinematic, data, data video, dataviz, datavis, visualization, news, newsgames, immersive, mixed reality, 3d}, and \textit{video}. We searched the portfolios of the creators of the visualizations found initially and their collaborators. A full list of the cinematic visualizations can be seen in Figure~\ref{fig:table} in the appendix of this paper; we refer to these studies by identifiers throughout the paper (e.g., CV4 refers to the fourth example in the table). Our analysis considered 50 cinematic visualizations. While the corpus is not exhaustive, the examples expose the variety of media (interactive news articles, YouTube videos, and TV segments) which cinematic visualizations occupy and the messages that they deliver.
The examples visualized a broad range of data types, including datasets both with and without physical and geographic dimensions.

% We call these \textbf{cinematic visualizations}:  narrative visualizations that embed data into a three-dimensional, time varying scene, employing one or more cameras to direct the relationship between a viewer and the scene to tell a highly visual type of data-driven story. 

% \matt{say something more about why this generalizes to all cinematic viz}.

% Cinematic visualizations embed data into a three-dimensional, time varying scene, and draw heavily on techniques from cinema to tell a highly visual type of data-driven story. While these visualizations have overlap with existing genres of narrative visualization (e.g. data videos~\cite{segel2010narrative}), we embrace Segel \& Heer's position that genres are not mutually exclusive. \jeff{This paragraph feels a bit redundant with prior sections}

% We present a detailed analysis of three of the cinematic data visualizations. We choose to highlight these three visualizations because they collectively employ a broad range of the stylistic techniques that we observed, and portray exemplary narrative applications of them. 

% ; however, cinematic visualizations can be expensive and time consuming to create. 

Rather than empirically evaluate specific design patterns utilized in the visualizations, we turn to the means of understanding plot devices~\cite{segel2010narrative}, sequencing~\cite{amini2015understanding}, and film style~\cite{bordwell1997film}. We analyzed the style of each example along the dimensions of \textit{mise-en-sc\`ene}, \textit{cinematography}, \textit{editing}, and \textit{sound} using the 4-step analysis process described by Bordwell and Thompson~\cite{bordwell1997film}, a canonical method of film analysis. For each example we first identified the main communicative goals of the visualizations, and then studied the salient techniques applied within the mise-en-sc\`ene, cinematography, editing, and sound which supported these narrative goals. We then used iterative coding to categorize the salient techniques used across the examples. Usage of these techniques are shown in Figure~\ref{fig:table}, for example we recorded many ways in which a viewer's attention is guided (through color, light, annotations, and narrators in the mise-en-sc\`ene) and use of cinematographic techniques like point-of-view perspective and user-controlled cameras. The table shows that the medium of the cinematic visualization has some impact on the techniques used, for example cinematic visualizations embedded in online articles rarely use sound, but often utilize user-paced segments, while those presented as videos make heavy use of sound.

\subsection{Design Techniques}

Through this analysis we identified salient recurring techniques that were frequently applied to support the communicative goals of the visualization, including the use of in-situ narrators, anthropocentric perspective, resolution of scale, and story-driven cameras. 

% \begin{figure}[h!]
%   \vspace{-28pt}
%   \caption{A picture of a gull.}
%   \centering
%   \includegraphics[width=\linewidth]{figures/in-situ.png}
% \end{figure}

\textbf{In-situ narrators mediate interactions with diegetic data.} Perhaps the most novel technique that we identified in cinematic visualizations is the use of in-situ narrators, in which the mise-en-sc\`ene contains a character that interacts directly with on-screen, \textit{diegetic data}.\footnote{Something which is diegetic exists in the same universe as the characters on screen; we use the phrase \textit{diegetic data} to refer to data-driven elements which are part of---rather than composited over---the scene shown in the frame.}
% In contrast to traditional cinema in which the camera plays the role of an unseen observer, cinematic visualizations frequently include narrators who speak directly to the camera, breaking the ``fourth wall''~\cite{bell2008theories}. 
In contrast to traditional documentary visualization narrators who might participate from off-screen (``voice of god'') or refer to data visualizations rendered as two-dimensional holograms or composited over top the video~\cite{bradbury2020documentary}, \textit{in-situ narrators} are understood by the viewer to be able to see and interact with the diegetic data either through the use of superimposed data visceralizations (CV35, 40, 42, 43) or, in one case, data physicalization~\cite{jansen2015opportunities} (CV41).
% Characters act as if they are  
This (typically) mixed reality environment serves an important role for narrative visualization, allowing the on-screen narrator to mediate interactions between the audience and the graphics, letting them provide additional context and push the storyline forward. These narrators, essential components of the mise-en-sc\`ene, can also help concretize a visualization's anthropocentric perspective, reinforcing the idea that data is being displayed at a human scale.

In \textit{The Dangers of Storm Surge} (CV42), one exemplar of this technique (Figure~\ref{fig:weather}) produced by the Weather Channel, a news anchor wearing a blue jacket explains the dangers associated with flooding due to storm surge. 
%   The video starts with a close up of the anchor before panning out to reveal computer-generated graphics showing various levels of flooding. 
  The graphics are coordinated with the narrator's script and appear to respond to his dialogue,  
  the composition of the frame inviting comparison between the man and the height of the water. 
The narrator is the primary subject from the start of the clip, positioned centrally in frame and maintaining focus due to visual cues like his bright blue coat, the circular platform upon which he stands, and the shot composition.
To call attention to the water's height at certain key moments, a brightly colored annotation is projected onto the crest of the surge.

% Although mixed reality is not strictly necessary to enable mediated interaction by a narrator (see CV37,39 for examples where mixed reality isn't used), it was used to create visually striking examples.

\textbf{An anthropocentric perspective transports viewers and enables drama.} One notable aspect of cinema is how the camera is able to transport the audience into the scene: people watching suspend disbelief~\cite{ferri2007willing} to allow themselves to wholeheartedly imagine, or ``believe'', that they are in the scene, seeing things through the camera lens. 
That is, the camera's perspective becomes the viewer's point of view, they are one and the same. 
% This anthropocentric perspective is created through careful placement and control of the camera. 
The height, angle, and distance of a camera in relation to objects in the scene all play a role in how a viewer will interpret and respond to the frame that they ultimately see~\cite{bordwell1997film}. When a camera is placed high above a setting, the viewer feels like they are also high above it. When a camera is placed at eye level, a viewer feels as if they are standing there watching the subject. For example, both CV1 and CV26 utilize unit visualizations and concrete scales to visualize quantities in relation to the size of Manhattan, but each uses perspective to impact the viewer's experience in a different way. In CV1 the data being displayed (plastic bottle usage) is not directly related to the locations being used as concrete scale referents, and an overview shot is used, letting the viewer absorb the scale of the data rather than the details and textures of the city itself. In contrast, CV26 begins with a shot from a camera placed at eye-level, looking at several of the city's ubiquitous yellow taxis, transporting viewers to the city at street level, and forcing them to reckon with the data being displayed (New York City's annual green house gas emissions) in a much more visceral way~\cite{lee2020data}.

% \begin{figure}[!b]
%   \vspace{-24pt}
%   \caption{A picture of a gull.}
%   \centering
%   \includegraphics[width=\linewidth]{figures/anthropocentric.png}
% \end{figure}

Some cinematic visualizations place the camera perspective somewhere that is humanly impossible. However, if the audience suspends disbelief, the camera can carry the viewer through these otherwise inaccessible spaces, for example, CV12 shows an animation of the Cassini spacecraft as it orbited and eventually crashed into Saturn.
\textit{Choice and Chance} (CV11), visualizes the events of the 2016 Pulse night club shooting in Tampa Bay, positions a camera looking ``through'' the roof of a nightclub. Because the scene is shot using a digital model instead of a real location, the roof of the club can simply be removed and problems of occlusion go away. Changing perspectives can also shift the subject of the scene or add emotional content, for example, when the camera moves to reveal something that wasn't already in the frame, the audience experiences seeing it for the first time. In \textit{Choice and Chance} the camera moves to different vantage points throughout the model as the story progresses. The camera remains in an overview shot for the majority of the article, but moves to ground level at the climax, elevating the intensity of the shot by placing the viewer into the perspective of a bystander.

% \begin{figure}[b!]
%   \vspace{-24pt}
%   \caption{A picture of a gull.}
%   \centering
%   \includegraphics[width=\linewidth]{figures/in-situ.png}
% \end{figure}

 \textbf{Author-defined camera trajectories can be played, paused, and (lightly) modified by viewers.} The cinematic visualizations that we analyzed tended to use author-driven narrative structures~\cite{segel2010narrative}, with most user interactions consisting of the user clicking or scrolling to trigger the visualization to continue to the next stage (e.g., CV2, 5-17, 21-22). Operationally, this requires animating the position and orientation of a digital camera model along a track specified by the author, and has been used heavily by cinematic visualizations embedded in articles (16 out of 22). The other way in which (constrained) interactivity was employed was allowing the manipulation of 3D models. In most cases this means the user can position the camera at a particular location around the model (see CV17 for a stereotypical example). These models might be scientific (CV13,17) or cultural (CV5) objects that would be otherwise inaccessible to the audience viewing the visualization. It is common for orbital cameras to be used, constraining the camera's focus to remain on a particular object of interest while allowing the user to exercise control over viewing angle and zoom level (Fig.~\ref{fig:table}D). Cinematic visualizations that support these interactions must be rendered in real-time, limiting the fidelity at which the models may be rendered.

% \textbf{Expressing Visualization Techniques}
\textbf{Visualization techniques are combined toward resolution of scale.} While we traditionally think of 3D graphics as ineffective for encoding quantities~\cite{cleveland1984graphical}, a recurring theme in our examples is the use of 3D graphics to visualize and communicate quantities of a massive scale (e.g., CV1, 6, 8, 26-28). 
Quantities at a scale beyond what we experience in daily life (i.e. \textit{hyperobjects}~\cite{morton2013hyperobjects}), like amount of carbon dioxide emitted from NYC annually (CV26), may be especially difficult for people to picture because we rarely, if ever, interact with quantities of such a size.
Cinematic visualizations can convey a quantity of scale in a concrete and affecting way by using cinematography to establish the viewer's point of view from the ground, a position which often serves as the implicit zero point of a y-axis.
% of a concrete scale~\cite{chevalier2013using}.  
We observed that several visualization techniques are naturally expressed in cinematic visualizations, including data visceralizations~\cite{lee2020data}, unit visualization~\cite{park2017atom} and concrete scales~\cite{chevalier2013using}. For example, in CV27 the viewer sees a city park, including trees, people standing in a grassy field, and a ten meter tall blue sphere representing the actual size of one metric ton of CO\textsubscript{2} (\textit{data visceralization}). As the scene progresses, many more spheres appear, each representing one metric ton of CO\textsubscript{2} (\textit{unit visualization}), until so many appear that the camera must zoom out, above the park, observing the growing pile of spheres in comparison to the city buildings  (\textit{concrete scale}).

% \begin{figure}[t!]
%   \vspace{-24pt}
%   \caption{A picture of a gull.}
%   \centering
%   \includegraphics[width=0.5\textwidth]{figures/in-situ.png}
% \end{figure}

\begin{figure*}[ht!]
  \centering
  \includegraphics[width=\linewidth]{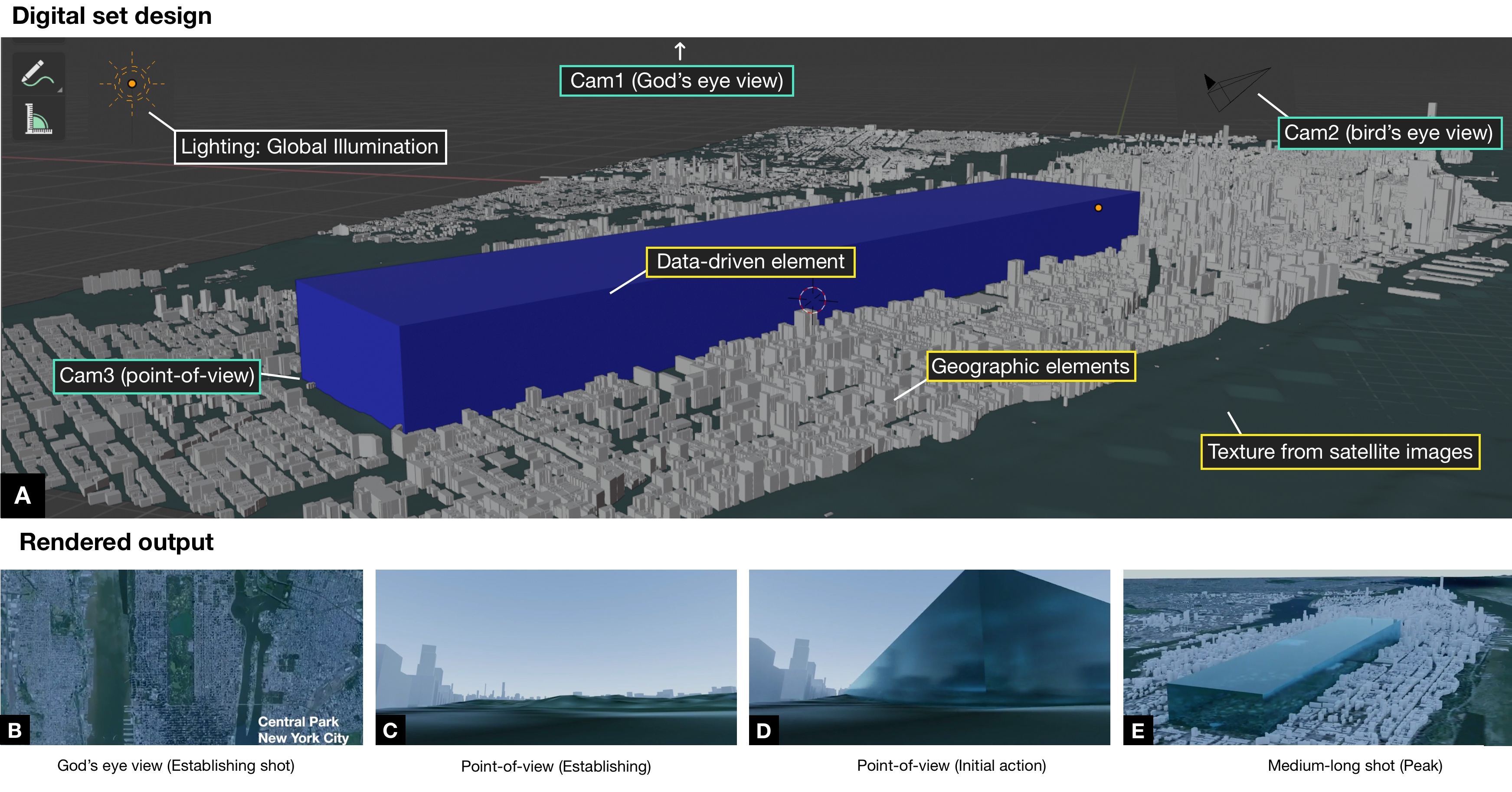}
  \vspace{-18pt}
  \caption{\textit{How Much is a Gigatonne?} shows one gigatonne of ice in Central Park, New York. A digital set (A) is designed including multiple cameras, lighting, and data-driven and contextual elements. Footage from the various cameras is composed to create the final sequence (B-E). This was one of several videos that we developed for an article published on NASA's climate website. View the full videos at  \url{https://cinematic-visualization.github.io/}.
%   The ice block is over 1,100 feet tall, comparable to the height of many of the city's buildings.
}
  \label{fig:gigatonne}
  \vspace{-10pt}
\end{figure*}

Objects which are used as backdrops---for example a city skyline (CV11) or parked car (CV42, Fig.~\ref{fig:weather})---may serve double duty as concrete scale referents and contextual elements. The use of 3D graphics affords designers the ability to use concrete scales (CV1, 26) and visual analogies (CV29, 36) to (re-)contextualize the size of objects, and digital sets are  constructed to facilitate comparisons that are impossible to make directly in the physical world (CV1, 27) and use point-of-view perspective to impart a visceral sense of magnitude. 
% \textbf{Unit Visualizations, Visual Analogies, \& Concrete Scales} 
The visual medium is rich with possibilities for analogy. 
% In some cases, the analogy is to transform scale. 
% Shown in Figure~\ref{fig:mixed-reality}, 
For example, in \textit{[REALISTIC] Elephant rocket fuel - Saturn V} (CV29, Fig.~\ref{fig:scale}), designer Maxim Sachs renders the launch of the Saturn V rocket, except that the rocket expels elephants behind it as it travels, rather than exhaust. The elephants represent the mass of fuel that is being expended. By juxtaposing these images, Sachs is able to re-frame an abstract quantity of rocket fuel in terms that people may have more familiarity with, and do it with a sense of humor that may make the visualization overall more memorable or engaging for its audience~\cite{borkin2013makes}. In a more  typical case, the narrator of CV40 asks the audience to imagine if Earth were the size of a tennis ball, and then, using this new scale, shows the relative size of different planets, moons, and stars. These planets are compared against one another, rendered into real-world footage including a narrator who provides guidance and relevant facts about the celestial objects. They are shown embedded into several settings, for example an office, a Los Angeles street, and the New York City skyline. 

\subsection{Constraints}

\textbf{The time-based format does not support a high data density.} Traditional information graphics often present a data-dense display with minimal ``non-data ink''~\cite{tufte2001visual} to remove possible distractions and optimize the display for tasks such as value look-up and comparison. In some cases, designers may choose to add additional illustrative features to increase the memorability of the visualization~\cite{borkin2015beyond}. In contrast, cinematic visualizations utilize diegetic data, embedded in a three dimensional scene with other elements which contextualize the scene (see CV35 for a striking example). In cinematic visualizations (e.g. CV40,42) the elements surrounding the data fulfill a dual role as both data and non-data ink; they add spatial presence to the visualization~\cite{cairns2014immersion},  supporting a sense of transportation to the virtual world for viewers, while simultaneously serving as guides and axes, points of reference for concrete scales~\cite{chevalier2013using}. Rather than densely packing data, we see that cinematic visualizations often only show one or a few data points in the frame, favoring to include additional contextual elements that help add emotional resonance to the data-story being told.
% , and enabling an author-driven~\cite{segel2010narrative} path through the data set. 

\begin{figure*}[ht!]
  \centering
  \includegraphics[width=\linewidth]{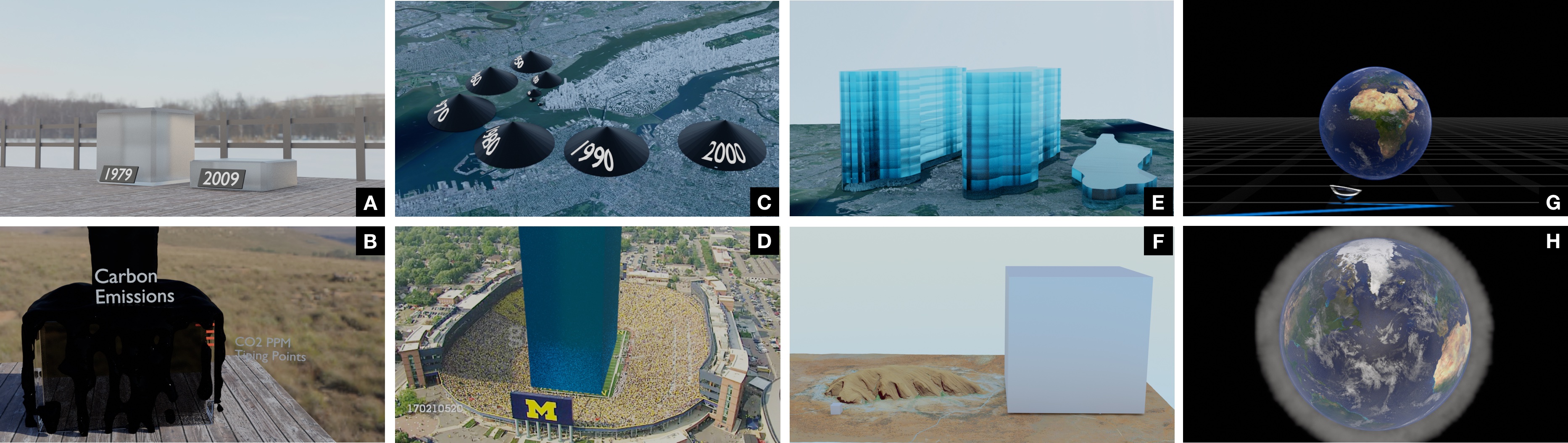}
  \vspace{-18pt}
  \caption{We explored many different designs, these were left on the cutting room floor. The designs were dropped for reasons including poor perceptual effectiveness (A-C), locations too small for the scale of the data (D-F), and designs too illustrative and not physically accurate enough (G-H). It was particularly difficult to identify locations that were broadly recognizable from a 3D reconstruction but also suitable to server as a concrete scale referent.}
  \label{fig:graveyard}
  \vspace{-10pt}
\end{figure*}

\textbf{Designers trade-off between perceptual effectiveness and dramatic narrative.} Visualizations that employ 3D graphics are often ineffective perceptually. These graphics may use sub-optimal encoding channels like volume and can further bias judgement through distortion and occlusion. Cinematic visualizations are not appropriate when the task is centered around value judgements. Instead, we see cinematic visualizations effectively used when a rough estimate of values is sufficient and the precise value is not of central importance (e.g. CV29). 
Many of the cinematic visualizations that we analyzed use a volume encoding to display data (CV1,6,26,27,35). Volume is a less effective encoding channel compared to position and may cause the audience to misestimate the true quantity. This trade-off may be acceptable depending on the data being presented and the precision with which the author hopes it will be apprehended.

\section{Case Study: \textit{How Much is a Gigatonne?}}

We collected and studied the aforementioned cinematic visualizations while exploring designs to support the communication objectives of NASA's Earth Science Communications Team. Climate change is a complex, multi-faceted issue of global importance~\cite{pachauri2014climate} and the team is tasked with maintaining \url{climate.nasa.gov}, a website that tracks vital statistics about Earth's climate, and delivers updates about global warming to a diverse global audience of millions of readers. The team uses traditional information graphics~\cite{change2017vital}, as well as narrative visualizations (e.g.,~\cite{forgedbyfire2019}), to highlight how scientists know that anthropogenic global warming is truly happening, what changes have taken place in Earth's climate so far, and why it is an important topic for readers to understand even if it does not seem to be affecting them. However, the team sought  data-driven stories that more viscerally engaged their audience and connect the planetary scale data of climate change to a human scale that readers can readily understand.
% We studied cinematic visualization techniques that strongly support these goals with the intention of using what we learned to develop new visualizations to be published on NASA's climate site.

 Within the domain of climate change communication is a range of research investigating how to effectively communicate the latest science to a broad audience.  High level principles of climate change communication have been synthesized by the Center for Research on Environmental Decisions~\cite{shome2009psychology}. We think cinematic visualizations are well suited to satisfy principles  ``Get Your Audience's Attention`` and ``Translate Scientific Data Into Concrete Experience.'' Here we describe how our work creates connections between ongoing investigations in narrative visualization, computer graphics, and film art to achieve this. 

Guided by editorial priorities set by NASA's Earth Science Communication team, we produced an article consisting of a several cinematic visualizations to communicate massive quantities related to climate change. We endeavoured to make them interpretable and meaningful to a broad public audience. These visualizations were eventually published to an audience of millions.
% Our first visualization communicates the size of one gigatonne of ice. 
Here we describe our design process to create cinematic visualizations, identifying a general workflow of use to practitioners who wish to create this type of visualization themselves, and to tool-builders who wish to provide better support for authoring cinematic visualizations in the future. As with visualization production in general, these steps are not necessarily linear; rather, the process is iterative and error prone, and may require going back to earlier steps if it becomes apparent that a design is not working. We experienced many failed attempts (see Figure~\ref{fig:graveyard}) before arriving at our final designs.

\subsection{Pre-Production}

% Here we work with domain experts (communications professionals tasks with informing the public about climate change). We decided that it would be interesting to show a series of quantities, starting with one gigatonne of ice. 
\textbf{Narrative.} Quantities of ice loss are measured in gigatonnes, a unit of mass corresponding to one million metric tons. Statistics about ice loss are often reported using this unit, for example Earth's polar ice caps are losing about 426 gigatonnes of ice per year, at the time of writing. The scale of the unit here hides the fact that 426 gigatonnes is a massive amount of ice. Our goal was to provide a visualization that would allow our audience to better interpret these statistics going forward. We collected statistics on ice loss in Greenland and Antarctica (the two ice sheets) over the course of significant periods, such as the amount of ice lost between 2002-2017 when NASA's Grace satellite was actively observing the polar ice caps, or since the start of the 20th century (5,000 and 49,000 gigatonnes, respectively).

We settled on cinematic visualization because it is a natural fit for the use of concrete scales, we wanted to draw people’s attention, there is a relatively small amount of data that we are showing, and we wanted to display the data in a context that conveyed corporeal urgency. Given the affordances identified in Section 3, a cinematic visualization was an appropriate choice for our task of visualizing quantities related to climate change in a way that would capture the attention of our audience and allow them to comprehend the data in a concrete way. We ultimately chose the form factor for our visualization to be an interactive article containing a series of short cinematic visualizations. The visualizations were embedded as pre-rendered videos, which could be loaded dynamically, allowing for a certain amount of interactivity. Depending on the use case, one must determine whether real-time rendering is needed or not. Using real-time rendering limits the level of \textit{photorealism}~\cite{pharr2016physically}, but enables another level of interactivity, letting the user control the camera and interact with elements in the scene (Fig.~\ref{fig:table}D). We intended the narrative structure of our visualization to be largely author-driven~\cite{segel2010narrative}, and decided that real-time rendering was not required. 

After determining that a cinematic visualization was appropriate, we began outlining possible scripts and creating storyboards in which we sketched ideas for locations, cinematography, and sequencing of shots. We first sought to identify locations that would serve as effective backdrops, allowing people to gain a concrete understanding of the size of data in familiar locations. We considered natural locations like the Grand Canyon, Monument Valley, Mt. Everest, and Uluru, urban environments like Houston, New York City, San Francisco, and St. Louis, and other man-made sites like football stadiums and the Hoover Dam. Within each of these environments we created sketches to help determine the camera placement, mise-en-sc\`ene, data, and annotations that the visualizations would require, and wrote rough scripts to define the narrative structure. 

While we wanted to place data in a variety of different environments so that our diverse audience would be able to connect, ultimately many of these locations were not used. See Figure~\ref{fig:graveyard} for examples of some of the locations that were not able to support both focus and context at an anthropocentric perspective. The final article consisted of videos visualizing one, then 5,000, then 49,000 gigatonnes of ice. The videos were embedded throughout the text of an article which provided context. In the first and last videos the user could click to choose to play videos displaying the relevant quantity of ice in different locations. Here we look closely at the design process for the first video, showing one gigatonne of ice in Central Park, New York City.

% \clearpage

\subsection{Principal Photography}

% \begin{figure}
%   \includegraphics[width=\linewidth]{figures/gigatonne.png}
%   \caption{Scenes from \textit{How Much is a Gigatonne?} showing one gigatonne of ice in Central Park, New York.}
%   \label{fig:gigatonne}
% \end{figure}

With the storyboards and scripts ready, the source footage that would make up the final video needed to be created. We chose to use Blender for this process, which provides both an interactive GUI-based interface as well as a Python API that allowed us to load, transform, and bind data to objects in a 3D scene. We created renders for many different scenes, although ultimately ended up using a small number of them in our published pieces.

\textbf{Mise-en-sc\`ene.} The elements that constitute the mise-en-sc\`ene of a cinematic visualization need to be created and arranged. Because many of our scenes take place in real-world locations, we were able to utilize existing open data sets to import geographic data, including 3D models of buildings and terrain data. In addition to elements derived from real-world locations, we added elements which would be parameterized by data, for example the large block of ice placed in Central Park (Figure~\ref{fig:gigatonne}). After the models have been created, they need to be assigned a material, which (along with lighting) will determine how they appear in final renders. We chose to use a flat shading for the buildings and other environmental elements. This gave these elements less visual weight while still allowing them to be easily identifiable. We considered using a similar flat style for the data elements, but ultimately decided to add a more photorealistic ice material which would allow the data to stand out against the buildings and reinforce the idea that we were showing a concrete amount of ice. While many of the examples that we saw utilize a studio lighting setup to control shadows and reflection, we opted to use simple global illumination to emulate the sun shining in our outdoor scene. This meant our lighting was realistic for the location and the setup was quite simple, but we were limited in our ability to use lighting as a tool to guide attention, as we saw it used (for example) in CV15.

% \textbf{Data Binding.} 
With the scene constructed, the next step was to bind the data. This was the point at which we realized that many of the set locations were not going to work with the data we were hoping to visualize (``data changes everything''~\cite{walny2019data}). For example, a gigatonne of ice placed in a football stadium (Fig.~\ref{fig:graveyard}D) would extend over 200 kilometers into the sky, making it difficult to view both the diegetic data and the stadium itself simultaneously.  
% \matt{Add specifics about why at least one didn't work. Reference Figure~\ref{fig:graveyard} D or F.} 
For our visualizations we were simply assigning the dimensions of a primitive 3D object based on calculations related to the mass of ice melt over specific periods, along with the density of ice, in order to create blocks of ice which were physically representative of the quantity lost.

\textbf{Cinematography.} After we incorporated our data into the scene it was time to add animation and cinematography. Blender supports a keyframe-based animation system which made it simple to add basic animations to the size and locations of elements in the scene, as well as the position and perspective of cameras. Working off of the storyboards that we had created, we placed cameras (shown in Figure~\ref{fig:gigatonne}) that would be physically realistic and familiar: we use three cameras, one a human point-of-view, one a bird's eye view (as if it were taken from a helicopter circling the city), and one a "god's eye view" taken from the perspective of a satellite overhead. The satellite camera allowed us to create an initial establishing shot, while the other cameras provided views that supported a ground-level view as well as an overview. When sequenced together, these camera perspectives allow us to present focus plus context~\cite{card1999readings} to the viewer, and support our narrative goals~\cite{amini2015understanding}. 

% Settings chosen to map onto human biology. Camera settings: 
% aspect ratio, - why do we prefer widescreen to vertical video? Because our screens are wide? I don’t really have a good answer to this.
% Focal length - settings impact what is visible in the frame. A low focal length is useful for wide-angle shots, compared to a high focal length which can be used for telephoto lens (close ups). A focal length can be used that matches that of human vision to reinforce the human centered perspective. Depth of field - useful for guiding attention (shallow depth of field means only part of image is in focus). Camera position and orientation~\matt{Need to expand...}.

\subsection{Post-Production}

Once the source material was created, we needed to edit it to form a coherent narrative, for example by combining multiple videos in sequence, adding annotations on top of the video to add context, and adding sound to add presence, guide attention, and provide details. Any visual effects must be added at this stage. For example, in the case of embedding digital data objects into physical footage of a narrator, a ``match moving'' process to align the digital and physical scenes would need to be performed~\cite{dobbert2006matchmoving}.

\textbf{Editing.} We combined footage from multiple cameras, composing shots into a narrative structure, starting with establishing shots, then initial action, peak, and finally release~\cite{amini2015understanding}.  The sequence of images is important to advance the role of narrative, pacing, and mood.
% \textbf{Annotation.}
Narrative visualizations often include annotations to provide additional context and explain to viewers what it is they are seeing. In the case of cinematic visualizations these annotations can be composited over the source footage using standard video editing software. Some examples that we saw embed annotations directly into the 3D scene itself, which requires them to be embedded in the source footage directly. We chose to composite annotations rather than include them ``in-situ'' as it facilitated more rapid iteration during the editing process, allowing us to change the timing, location, and content of annotations, without needing to re-render any of the source footage\,---\,a potentially time-consuming process. 

\textbf{Sound.} 
In our work we ultimately did not use audio, instead opting to embed the videos in a larger text article, which would serve to provide viewers with context for the visualization. This is a limitation and something to be explored more in future work, as audio can be a useful tool in cinematic visualization to set tone and drive narrative.

\subsection{Publication} Once the article was completed and approved for publication, it was posted to NASA's climate website.  We did not collect detailed metrics on how readers interacted with the videos on the article itself, but can see how users responded to posts on the NASA Climate Facebook, Instagram, and Twitter pages. These posts---which contained a link to the article and (in some cases) directly embedded the video set in New York City---were collectively viewed tens of thousands of times, received thousands of engagements (likes, comments, shares), and the article was subsequently shared by other organizations such as the United States Department of Agriculture and the World Meteorological Organization, as well as by individual scientists and meteorologists. 

Across all of the social platforms users left 94 direct comments, with topics ranging from positive (for example, some explicitly expressing that they like this type of visualization ``We need more of these types of comparisons in the media'', ``This is an amazing visualization. Thanks NASA!'', or asking for similar visualizations of different quantities ``It would be very interesting to see this illustration but with the predicted sea level after all the ice in Greenland and Antarctica melt. Can you show that?'') to concern about the data being visualized (``Oh my God. Come to our aid.'', ``Thanks for helping us comprehend the enormity of this sad news!'', \textit{a GIF of a cartoon rodent crying}) to climate change denial (``Where's your proof'', ``Wow, as much as 2 millimetres. Measured by satellite too''). The comments were distributed roughly uniformly across the three types (\textit{positive attitude toward visualization}, \textit{concern about climate change}, and \textit{climate change denial}), but varied heavily across platforms, with users on Facebook expressing concern or denying that there is a climate problem, users on Instagram leaving both positive and concerned comments, and users on Twitter expressing a range of concern,  denial, and a positive attitude toward the graphic.

\section{Discussion}

% and then present a set of design guidelines that we synthesized through our analysis of existing cinematic visualizations and real-world experience creating a series of cinematic visualizations for NASA. 

Cinematic visualizations can engage viewers with dramatic and visceral presentations of data, highlighting particularly important data points, and presenting an author-guided tour through data embedded in a relevant context. On the other hand, they may be poor choices for communicating large amounts of data and are not optimal in terms of perceptual effectiveness. If a cinematic visualization is appropriate, it will require a broad range of skills\,---\,such as cinematography, narrative, 3D modeling, video editing, and possibly acting\,---\,and a time-consuming iterative design process. 

% Before embarking to create a cinematic visualization, designers should consider whether the form is appropriate, and if they have the skills and support needed to create one. \matt{use more appropriate tone for previous sentence. There are entire professions dedicated to each of the points we are mentioning; this is going to be a summary job.}

\subsection{Challenges of Creating Cinematic Visualizations}

While cinematic visualizations can capture the attention of their audience and help viewers relate to the data in a concrete way, they can be challenging and time-consuming to produce. Here we discuss some of the challenges inherent in creating an effective cinematic visualization. 
% There are a number of broad challenges that designers will face when crafting a cinematic visualization. 
% Here we concentrate on issues that are specific to this particular genre, rather than challenges of narrative visualization more generally.

One of the most apparent difficulties of cinematic visualization is the potentially overwhelming size of the design space. Works in this genre typically use three visual dimensions, plus time and sound. The methods that allow us to analyze and critique cinematic visualizations (e.g.,~\cite{bordwell1997film}) do not necessarily help us to create them. That is, they are difficult to use generatively. While information designers are familiar with the attention to detail that is required when placing objects in a frame in order to achieve an effective visual hierarchy, in cinematic visualizations there are also objects outside of the frame that affect the style and tone of the visualization. For example, the placement of the camera in relation to the subjects, the focal length of the camera, and the placement and strength of light sources are all instrumental in creating a shot which can easily be decoded by viewers. 

There is a diversity of tasks that need to be completed in order to create a cinematic visualization, each requiring a separate set of skills. For example, in addition to skills required for traditional visualization (data analysis, transformation, and visualization) and narrative visualization (understanding audience, storytelling, graphic design), cinematic visualization will often make use of animation, cinematography, lighting, motion graphics, 3D modeling, sound design, video editing, and (sometimes) acting. The skills that make one a good 3D modeler are not necessarily the same skills that make one a good storyteller, and so graphics of this type often require a diverse team to create. Furthermore, for ray-tracing renderers, there is a large gap between prototypes and final rendered output, challenging the iterative design process.

\subsection{Considerations for Cinematic Visualization Creators}
% \matt{This section still needs to be updated to include more specific references.}
% The guidelines for creating successful cinematic visualization largely overlap with those for designing narrative guidelines, but how these are operationalized are very different.

While cinematic visualizations share many of the same design goals of more traditional narrative visualization (e.g., guide the viewers' attention), the way in which these goals are operationalized differ. Here we highlight ways that these design goals were operationalized across the four dimensions of style, both in our own work and in the examples analyzed. For a full breakdown of the techniques used in each example, see Figure~\ref{fig:table}. 

% One of the biggest challenges in creating cinematic visualizations is the overwhelming size of the design space. 

% and utilize rules of photography (e.g., the rule of thirds) in order to create an effective visual hierarchy. 

% Designers would be well served to study The history of cinema in order to understand that visual language, which has been established over the past 100+ years. 
% The ability to include annotations through voice-over narration, on-screen text, or in-situ narrators can be utilized to provide additional context to help viewers understand the data.

% Cinematic visualizations can be used to transport the viewer to a particular place and time, and can be used to create particular perspectives or scenarios. 

% Through our survey we saw that effective cinematic visualizations are used to engage audiences while adding context and perspective to data, and are often used to help viewers understand scale by making use of unit visualizations, visual analogies, and concrete scales. 
% They shouldn't be used when precise value judgements are a core task for users, and they should not be used as an excuse to construct gratuitous 3D charts. 

\textbf{Mise-en-sc\`ene.}
Objects' sizes, colors, shapes, textures, and placement in relation to one another can all be used create an effective visual hierarchy. For example, to guide a user's attention in a cinematic visualization, a designer might choose to use lighting to cast a glow around an object (CV11), or change the object's color (CV2, CV13) so that it stands out.  In \textit{How Much is a Gigatonne}, the ice's large size,  color, and shine draw a viewers attention to  it in contrast with the surrounding buildings, which are smaller, grayscale, and matte. The mise-en-sc\`ene is designed both to communicate information---including using narrators (CV42), diegetic data (CV35), and visual analogies (CV6)---and to add dramatic affect (e.g. CV11, CV40).
% Cinematic visualizations may have additional goals such as using drama to 

% designed for dramatic affect. In this case the  

\textbf{Cinematography.} Perspective can be used both to drive narrative and to set tone, as well as to provide focus plus context. The position (CV26), angle (CV28), or focus (CV2) of a camera can be modified so that the object becomes the focal point of the frame. To help narrow the large space of possible cinematic visualizations, and make effective use of the frame, designers of cinematic visualization may study how shots are composed and sequenced in films.
In \textit{How Much is a Gigatonne?}, we rendered footage from multiple cameras in order to create close-up, medium, and wide shots. Some cinematic visualizations enable limited user-control of the camera, for example letting the user trigger the next stage of animation (CV9) or rotate their perspective (CV13).  Often the camera needs to track a particular object in the scene (CV12). If this object is in motion you may need to set your camera to track it. Planning the path of the camera so that the object of interest is not occluded by other objects and so that motion is smooth and visually pleasing can be difficult. This may be done algorithmically~\cite{yeh2011efficient, christie2003constraint} or by hand. 

\textbf{Editing.} Putting the footage into a particular order progressively reveals information to convey the authors' intended message. Editors may use footage from one camera at one location (CV29), or multiple cameras at multiple locations (CV40).  The editing techniques used in data videos---particularly the use of \textit{establishing}, \textit{initial}, \textit{peak}, and \textit{release} shots---has been studied in more depth by Amini et al.~\cite{amini2015understanding}. Similar to movie makers, creators of cinematic visualizations may use the technique of storyboarding to prototype and communicate their scenes in a lo-fidelity form before endeavouring on the time intensive task of 3D modeling and rendering. In \textit{How Much is a Gigatonne} we use establishing shots to situate the viewer before initiating action from the perspective of the ground level (an anthropocentric perspective), before cutting to the vantage point of a helicopter, using the city skyline as a concrete scale.

% Create storyboards and write scripts; anticipate a slow feedback loop. 

\textbf{Sound.} Audio can set tone (CV25), cue attention (CV28), and impart additional details through narration on (CV40) or off-screen (CV45). 
Music (CV29) and ambient sound (CV26) can affect the tone of the visualization and add presence to the scene, for example CV29 uses combines techno music and a visual analogy of of the weight of rocket fuel (measured in elephants) to create a humorous juxtaposition which may make the visualization more approachable and less dry. CV26 uses diegetic sound (taxi cabs honking) to reinforce the anthropocentric perspective. In \textit{How Much is a Gigatonne} we did not use sound (neither did most of the other visualizations that we analyzed which used an ``article'' format), but effective use of both the visual and auditory  channels has been shown to lead to improved outcomes in multimedia learning contexts~\cite{low2005modality}.

% \textbf{Use visual and auditory cues to guide attention.} can be done through use of light, camera position, material (color), motion, annotation, or via a narrator or character. Objects in a scene should have a consistent visual aesthetic. If you must break the aesthetic, do it intentionally. Don’t make objects in the foreground photorealistic but objects in the background cartoonish. If you are constructing a photorealistic set, there should be “contextual” elements that add to the immersion, e.g. trees, shrubs. If you are using making something abstract, consider removing background or contextual items completely. 

% \subsection{Discussion} 

% We believe that this is a promising style of visualization that can help to inform the public of the science behind some of the most pressing global issues such as climate change. 

\subsection{Implications for Authoring Tools}

As cinematic visualization is a newly emerging genre, there is relatively little tool support to facilitate authoring of this type of visualization. Instead, creators turn to general purpose 3D software that was designed to support a breadth of use cases such as architectural design, modeling, and narrative animation. These tools, while powerful and expressive, may overwhelm users with complexity that is incidental to the task of creating a cinematic visualization. For example, objects are assigned materials which are powered by low-level shader code. One can not choose, e.g., between “realistic” or “cartoon” aesthetics but instead must compose low level shader components to achieve the desired look.

These tools do not support the basic building blocks of visualization, such as easily ingesting data and binding data values to objects in a scene. Instead, users must write custom scripts to handle any such task. The interfaces in general are multi-modal: most 3D modeling work is done directly through a GUI, but data-driven work needs to be done in code; shaders are described using a directed graph. Authors are forced to context switch between drastically different environments, arguably making it harder to iterate.

The task of 3D rendering can be computationally intensive. Depending on the output resolution, complexity of the scene, and computing power available, a short (30 seconds) animation could take several hours to render. There is a large gap between the fidelity of the final renders and what a designer sees while constructing the scene. This setup makes it important to create test renders frequently, but makes it hard to have a rapid feedback loop.

\subsection{Limitations of our Work}

% We acknowledge that there are clear limitations to our research contribution, even so we believe that our work will be of interest to the information visualization research community.
Our survey was limited to 50 examples, taken from a limited set of sources. While not exhaustive, the examples implement a range of design techniques across a variety of applications. We do not provide an empirical evaluation of the work surveyed, instead choosing to use techniques of film criticism in order to analyze patterns used and identify the communication intentions of their producers. We similarly did not empirically evaluate our own work, and instead provide an account of our design process and detail our reasoning for important decisions that were made along the way.  Our work does not fully utilize the design space of cinematic visualizations that we identified; for example, we did not use sound at all, and all narration was done through written text with a few small overlays in the video. The experience might be improved by incorporating narration either on-screen or off~\cite{bradbury2020documentary}. 

\section{Conclusion}

We presented \textbf{cinematic visualization}, a genre of narrative visualization that uses techniques from cinema in order to enhance the presentation of data-driven stories.
A central contribution of this work is to identify a new \textit{genre} of narrative visualization that we then analyze in depth. The importance of genre is clear in other art forms like literature and cinema; however, it is invoked less often in the context of visualization research. We believe that this type of work is crucial for understanding the design of narrative visualizations, and thinking rigorously about how they can be constructed and deployed. While past work on narrative visualization looked specifically at the narrative structure, here we look at both narrative and style as formal systems that contribute to the dramatic experience of watching a cinematic visualization. To do this, we turned to theory from another form of art, film, in order to provide grounding in the features of style, and used analysis techniques established in that domain to deconstruct our case studies. 

We analyzed a variety of examples of cinematic visualization and the techniques that they employ towards certain narrative applications. Many of these visualizations show a relatively small amount of data (e.g., focusing on a single rate or quantity) as opposed to being data-dense. The non-data elements of the scene play an important role: they are used to set the location in which the shot is taking place and provide cues to viewers about where they are, what they are looking at, and why it is relevant. This approach is quite different from typical information visualizations, where data may be reduced to a minimal form, such as a line or a bar. Cinematic visualization instead tends to be more \textit{maximal} in its approach, such that the non-data ink is not reduced or omitted, but rather used to build up entire digital worlds around data points. This style encourages viewers to feel present in locations augmented with data objects, or to viscerally experience events that happened in the past, or are happening far away in the universe.

% \begin{figure}
%   \centering
%   \includegraphics[width=\linewidth]{figures/scientific.png}
%   \caption{In Resurrecting a Dragon a 3D model of a dinosaur fossil is presented. The viewer is directed through various aspects of the fossil via text as well as visual cues such as visual highlighting and camera movement.}
%   \label{fig:scientific}
% \end{figure}
 
% \subsection{Editorial v. Fidelity in Cinematic Visualization} 

% While cinematic visualizations do not strictly have to be three dimensional, many of the examples that we observed were.
Rendering data in 3D is a fraught endeavor, as the values being rendered can be obscured by humans' relatively poor ability to estimate and compare volume, and because the 3D projection can introduce distortion when trying to read values. Why would the creators choose to follow a cinematic path over one that more clearly and directly communicates the underlying data with precision? We argue that in choosing to treat a visualization as a cinematic experience, its authors might be looking beyond the immediate data, in order to viscerally ground that data in meaningful context. In other words, analytic precision is only one of several objectives that a visualization might help accomplish. In choosing 3D, we might diminish precision in service of other objectives.

% \jeff{Everything that follows in this paragraph is sufficiently detailed to belong earlier in the paper, rather than in the concluding remarks?}
% For instance in CV6 the goal of the graphic is to compare the size of two groups by comparing monopoly pieces that relate the two classes. In terms of precision, it can be very difficult for an observer to accurately compare the two values when displayed in this way. But some other, editorial value is introduced through the comparison. One such meaning might be that by placing the red pieces on top of the conical pile, the author makes the statement that the red hotels are situated at the top of the ``economic food pyramid.'' Another idea introduced by the particular editorial position is that by placing the pile of houses in front of the White House, they resemble the US National Christmas Tree. The juxtaposition of a religious symbol of equality, with data that characterize the influence of an elite, creates a tension between the symbol and its underlying data. These editorial elements provide a stronger signal than the volumetric comparison between the green and red houses, which actually require white text overlays to precisely describe.

\section*{Acknowledgements}

We would like to thank Susan Callery, Holly Shaftel, Randal Jackson, Daniel Bailey, Michael Gunson, Josh Willis, Joe Witte, and the Earth Science Communications Team at NASA's Jet Propulsion Laboratory for their support of this work. A portion of this research was carried out at the Jet Propulsion Laboratory, California Institute of Technology, under a contract with the National Aeronautics and Space Administration (80NM0018D0004).

\bibliographystyle{ACM-Reference-Format}
\bibliography{references}

\appendix

% \begin{landscape}
\begin{figure*}
  \centering
%   \vspace{-18pt}
    \includegraphics[width=\linewidth]{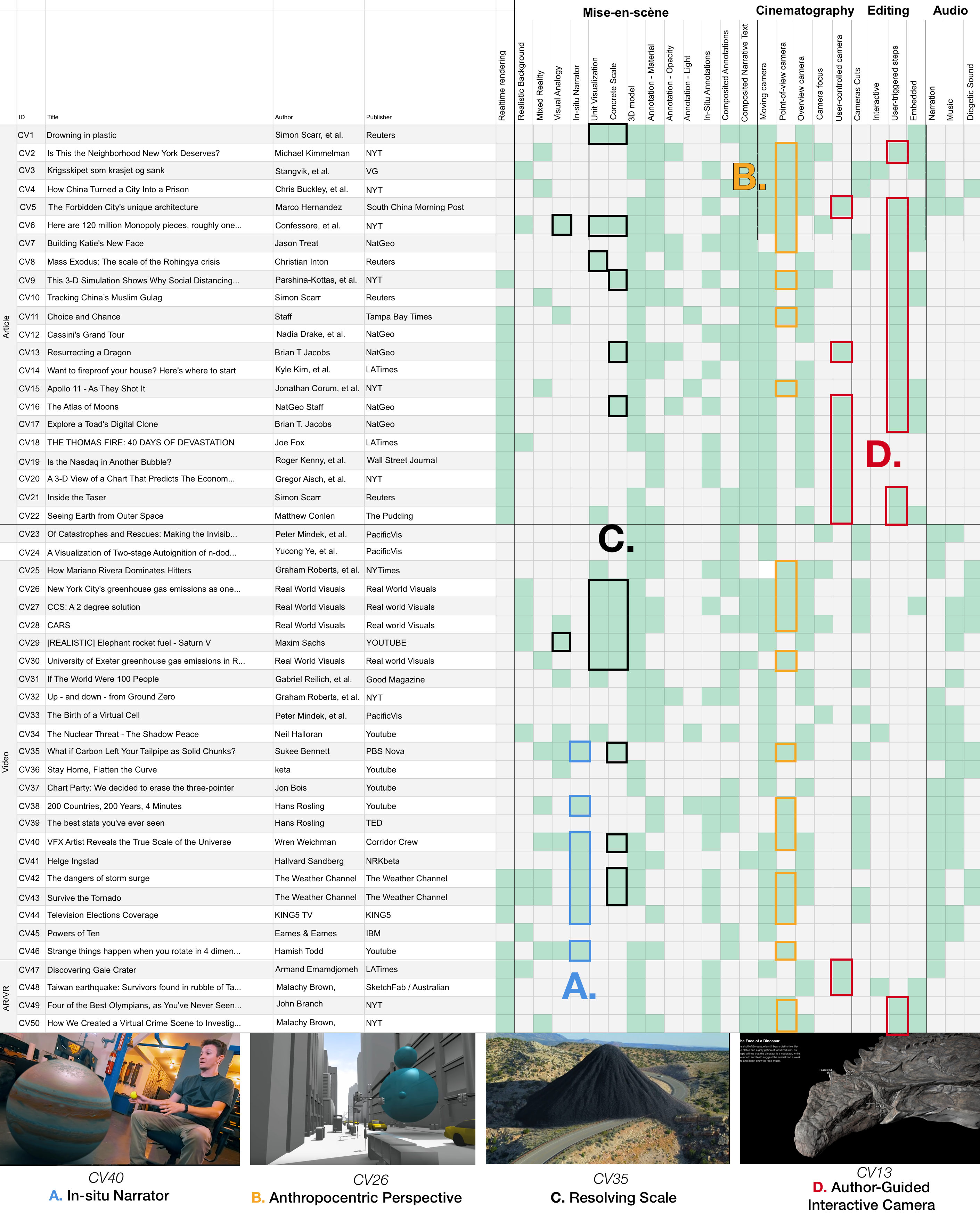}
    \caption{We analyzed the style of 50 cinematic visualizations using the features of mise-en-sc\`ene, cinematography, editing, and sound. 
    % We identified techniques across each of these dimensions of style, including (A) in-situ narrators, (B) anthropocentric perspective, (C) concrete scales, and author-guided, interactive cameras (D). 
    % Note that the use of techniques related to editing and sound varied greatly between cinematic visualizations embedded in online articles and those presented as videos. 
    An HTML version of this table including URLs for each row can be found at \url{https://cinematic-visualization.github.io/}.}
    \label{fig:table}
\end{figure*}
% \end{landscape}

\end{document}